\documentclass[conference]{IEEEtran}

\usepackage{cite}
\usepackage{amsmath,amssymb,amsfonts}
\usepackage{algorithmic}
\usepackage{graphicx}
\usepackage{textcomp}
\usepackage{xcolor}
\usepackage{hyperref}

\begin{document}

\title{R-LAM: Reproducibility-Constrained Large Action Models for Scientific Workflow Automation}

\author{
    \IEEEauthorblockN{Suriya Sureshkumar\textsuperscript{1,*}}
    \IEEEauthorblockA{\textsuperscript{1,*}Department of AI \& Data Science, RMK Engineering College, Chennai, India}
    \IEEEauthorblockA{240171.ad@rmkec.ac.in}
    \thanks{*Both authors contributed equally to this work.}
}

\maketitle

\begin{abstract}
Large Action Models (LAMs) extend large language models by enabling autonomous decision-making and tool execution, making them promising for automating scientific workflows. However, scientific workflows impose strict requirements on reproducibility, auditability, and deterministic execution, which are not satisfied by generic LLM-based agents. Unconstrained action generation can lead to silent state changes, non-deterministic executions, and irreproducible experimental results, limiting the applicability of LAMs in scientific settings.

In this paper, we propose R-LAM, a reproducibility-constrained framework for applying Large Action Models to scientific workflow automation. R-LAM introduces structured action schemas, deterministic execution policies, and explicit provenance tracking to ensure that every action and intermediate artifact is auditable and replayable. The framework supports failure-aware execution loops and controlled workflow forking, enabling iterative experimentation without compromising reproducibility.

We implement R-LAM as a lightweight Python framework and release it as an open-source PyPI package to facilitate reproducible research. An experimental evaluation of representative scientific workflows demonstrates that R-LAM improves reproducibility success rates and execution reliability compared to unconstrained LLM-based agents, while retaining adaptive control over workflow execution.
\end{abstract}


\section{Introduction}

Large language models (LLMs) have demonstrated strong capabilities in reasoning, code generation, and tool usage, enabling a new class of autonomous systems commonly referred to as agents or Large Action Models (LAMs). Unlike traditional LLMs, LAMs are designed to select and execute actions such as running code, invoking APIs, or orchestrating multi-step workflows. These capabilities have motivated growing interest in applying LAMs to automate scientific data analysis pipelines and experimental workflows.

Despite this promise, scientific workflows impose constraints that fundamentally differ from general-purpose automation. Scientific experiments are expected to be reproducible, auditable, and deterministic, such that results can be independently verified and extended by other researchers. Generic LLM-based agents often operate with implicit state, unconstrained action generation, and stochastic execution behavior, which can introduce silent state changes, non-deterministic outcomes, and incomplete provenance records.

To address these challenges, we introduce R-LAM, a reproducibility-constrained framework for applying Large Action Models to scientific workflow automation. R-LAM treats actions as structured, first-class entities, enforces deterministic execution policies, and records complete execution traces for provenance and replay. The framework enables adaptive experimentation through failure-aware execution and controlled workflow forking, while preserving reproducibility guarantees.

\textbf{Main contributions:}
\begin{enumerate}
\item \emph{Formal action schema}: A machine-readable specification of actions that decouples intent (what to execute) from implementation (how to execute it), enabling auditability without requiring knowledge of execution details.
\item \emph{Deterministic execution engine}: A mediation layer that enforces execution policies, isolation, and control without modifying the LAM's reasoning process.
\item \emph{Provenance-aware trace graph}: A complete execution history in DAG form that enables replay, fork-based experimentation, and failure auditing without re-execution.
\item \emph{Reference implementation}: A lightweight, open-source Python framework demonstrating that reproducibility constraints can be integrated into LAM execution without sacrificing adaptive control.
\end{enumerate}

\section{Challenges in Scientific Workflow Reproducibility}

Reproducibility remains a central challenge in computational and data-intensive scientific research. Prior studies show that many published computational results cannot be independently reproduced due to undocumented software environments, missing parameters, and incomplete provenance capture. Beaulieu-Jones \textit{et al.} demonstrate that even when code and data are available, substantial engineering effort is often required to reproduce results, motivating automated reproducibility mechanisms \cite{beaulieu2017reproducibility}.

Scientific workflows frequently contain hidden state and implicit dependencies that are not captured in workflow specifications or execution logs. Environment configurations, external tool versions, and ephemeral intermediate artifacts may influence results without being explicitly recorded. Suetake \textit{et al.} highlight the difficulty of validating workflow reproducibility in bioinformatics, even in containerized settings, due to the lack of automated equivalence verification mechanisms \cite{suetake2023workflow}.

Non-determinism further exacerbates reproducibility challenges. Stochastic algorithms, parallel execution, and hardware variability can cause repeated runs with identical inputs to yield divergent outputs. Horton \textit{et al.} emphasize that simple version control of scripts is insufficient for reproducibility, underscoring the need for comprehensive environment and provenance management \cite{horton2022reproducibility}. In contrast to business automation, scientific workflows must treat every execution as a first-class scientific artifact.

\section{Related Works}

\subsection{LLM Agents for Scientific Workflows}

Recent work explores using LLMs to assist scientific workflow development. Shin \textit{et al.} describe the evolution of scientific workflows in the agentic AI era, arguing that LLM-driven agents can progressively automate scientific processes \cite{shin2025agentic}. Gupta \textit{et al.} demonstrate that ChatGPT can assist users in understanding and authoring scientific workflows, though primarily in a human-in-the-loop setting \cite{gupta2023chatgpt}. FlowMind, proposed by Zeng \textit{et al.}, uses LLMs to generate structured workflows from high-level task descriptions \cite{zeng2024flowmind}. Masera \textit{et al.} introduce Snakemaker, which converts ad-hoc scripts into maintainable Snakemake workflows \cite{masera2025snakemaker}.

Beyond workflow generation, Large Action Models formalize agentic execution. Dodge \textit{et al.} define LAMs as LLM-based systems that generate and execute actions in external environments \cite{dodge2024lam}. Sangchai \textit{et al.} extend this paradigm to human-in-the-loop robotic systems \cite{sangchai2025robotlam}. Benchmarks such as AgentBench \cite{liu2023agentbench}, TheAgentCompany \cite{xu2025agentcompany}, and MedAgentBench \cite{jiang2025medagentbench} show that while LLM agents can perform multi-step reasoning, their execution behavior remains stochastic and difficult to control.

\subsection{Workflow Management Systems and Reproducibility}

Workflow management systems emphasize deterministic execution and provenance tracking. Beaulieu-Jones and Greene propose continuous analysis to automate reproducibility \cite{beaulieu2017reproducibility}. Mione \textit{et al.} present a DAG-based workflow system for reproducible self-driving laboratory experiments \cite{mione2024workflow}. Grayson \textit{et al.} study reproducibility at scale by re-executing Snakemake and nf-core workflows \cite{grayson2023snakemake}. WorkflowHub provides an open registry for sharing computational workflows \cite{goble2025workflowhub}. Domain-specific systems such as Atomate2 \cite{rich2025atomate2} and AlabOS \cite{fei2024alabos} further demonstrate reproducible workflow execution but rely on predefined pipelines without adaptive reasoning.

\subsection{Automated Experiment Orchestration and AutoML}

Automation of experimental design has also been explored. Gierisch and Mauerer introduce QEF, a framework for reproducible quantum software experiments \cite{gierisch2025qef}. Xu \textit{et al.} propose an AutoML-based workflow for design-of-experiments selection and benchmarking \cite{xu2024automl}. These systems automate parameter exploration but operate within fixed execution structures.

\subsection{Gap Analysis}

Existing approaches either emphasize autonomous reasoning without execution guarantees or deterministic workflows without adaptive control. Our work bridges this gap by treating reproducibility as a core constraint within LAM-based execution, enabling adaptive scientific workflows while preserving auditability, determinism, and replayability. Rather than extending a single prior system, our work synthesizes insights from LLM-based action models and reproducible workflow management to address execution-level gaps in scientific automation.

\section{Large Action Models: Opportunities \& Risks}

Large Action Models (LAMs) extend large language models by coupling high-level reasoning with the ability to execute concrete actions in external environments. This execution-centric paradigm introduces new opportunities for automating complex scientific workflows, but also exposes fundamental risks when applied to scientific settings where execution semantics, provenance, and determinism are critical. This section analyzes both aspects to motivate the need for reproducibility-aware execution constraints.

\subsection{Opportunities}

A central advantage of LAMs is their ability to support \emph{adaptive control}. Unlike static workflow engines, LAMs can condition future actions on intermediate results, enabling workflows that respond dynamically to observed outcomes. This allows scientific pipelines to adjust execution paths when unexpected data characteristics, quality issues, or partial failures are encountered.

LAMs also enable \emph{dynamic parameter tuning} during execution. Rather than fixing experimental parameters in advance, an LAM can iteratively refine configurations based on observed performance signals. This capability is particularly useful in exploratory scientific workflows, where optimal parameters are not known a priori and must be discovered through iterative experimentation.

Another key opportunity is \emph{failure-aware replanning}. Traditional workflow systems typically halt execution upon failure or require manual intervention. In contrast, LAMs can reason about failures, infer potential causes, and attempt alternative execution strategies, such as rerunning steps with modified inputs or substituting tools. This adaptivity has the potential to reduce human effort in managing long-running and complex scientific workflows.

\subsection{Risks}

Despite these advantages, the execution semantics of Large Action Models remain poorly defined in scientific settings. Because action selection is driven by probabilistic model outputs and executed through heterogeneous external tools, LAM-based systems lack inherent guarantees about state isolation, execution ordering, and outcome consistency.

One major risk is \emph{action hallucination}, where the model generates invalid, unsafe, or non-existent actions. Unlike hallucinated text, hallucinated actions can directly modify external state, corrupt datasets, or invalidate experimental results, often without immediate detection.

LAM-driven execution is also susceptible to \emph{non-deterministic behavior}. Variability in model outputs, stochastic decoding processes, and interactions with external systems can cause repeated executions with identical inputs to diverge. Such variability undermines the repeatability required for scientific validation and comparison.

Another critical concern is the presence of \emph{undocumented state changes}. LAMs may implicitly depend on or modify hidden state, such as temporary files, environment variables, or external service configurations, without explicitly recording these changes. These hidden dependencies complicate auditing, replay, and independent verification of results.

Collectively, these factors lead to \emph{result irreproducibility}, where outcomes cannot be reliably regenerated even when code and data are preserved.

\subsection{Implications for Scientific Automation}

Without explicit constraints, LAMs optimize for task completion rather than scientific validity. While this objective may be acceptable in general automation or business process settings, it is fundamentally misaligned with the requirements of scientific research, which demand determinism, auditability, and complete provenance. Addressing these risks is therefore a prerequisite for responsibly deploying LAMs in scientific workflow automation.

\section{Reproducibility-Constrained LAM Framework}

\begin{figure}[ht!]
    \centering
    \includegraphics[width=0.5\textwidth]{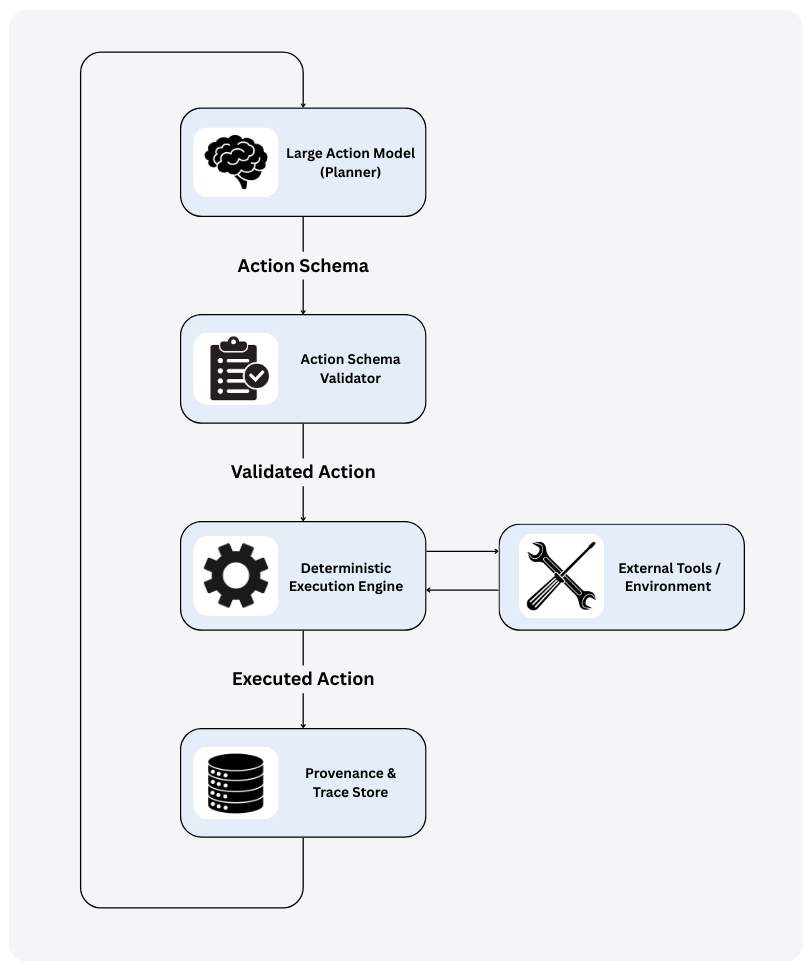}
    \caption{High-level architecture of R-LAM. The Large Action Model proposes structured actions that are validated and executed by a deterministic execution engine. All execution effects are recorded in a provenance-aware trace store, which provides observable state for subsequent reasoning. [Note: Final submission should include high-resolution PNG/PDF (>=300 dpi) per IEEE standards]}
    \label{R-LAM_HL_Architecture}
\end{figure}

This section presents the core technical contribution of this work: a reproducibility-constrained execution framework for Large Action Models (LAMs). The framework is designed to enable adaptive, agent-driven workflow execution while enforcing strict guarantees on auditability, determinism, and replayability. Rather than modifying the internal reasoning of the LAM, the framework constrains execution at the action level, treating reproducibility as a first-class system invariant.

\subsection{Action Schema}

At the foundation of the framework is a formal definition of an \emph{action}. An action represents the smallest unit of executable behavior issued by a LAM and is defined as a structured, immutable object:
\[
\begin{aligned}
a = \langle\;& id,\ type,\ inputs,\ parameters,\\
             & preconditions,\ effects,\ metadata \;\rangle
\end{aligned}
\]
where $id$ uniquely identifies the action instance, $type$ specifies the execution primitive (e.g., tool invocation, script execution), $inputs$ and $parameters$ define all required data and configuration values, $preconditions$ encode required state constraints, $effects$ describe expected state transitions, and $metadata$ captures contextual information such as timestamps, environment identifiers, and model configuration.

Actions are strictly declarative: they describe \emph{what} is to be executed, not \emph{how} execution is performed. This separation ensures that all executable intent is explicitly represented before execution, eliminating implicit or hidden behaviors.

\subsection{Execution Engine}

\begin{figure*}[t]
    \centering
    \includegraphics[width=1\textwidth]{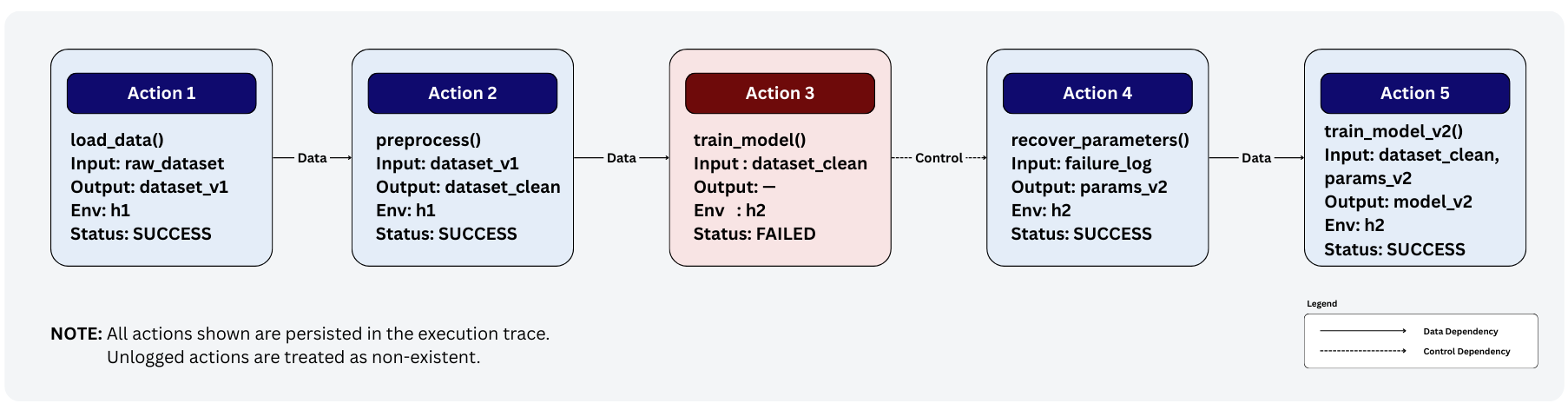}
    \caption{Execution trace graph structure. Each node represents a single executed action with its inputs, outputs, execution environment binding, and terminal status. Nodes marked with a failure indicator represent actions that terminated unsuccessfully. Critically, failed actions are retained in the trace (not removed or implicitly retried), enabling subsequent recovery actions to be explicitly linked. This design preserves complete execution history and ensures that all downstream results can be traced to observed events. [Note: Final submission should include high-resolution PNG/PDF (>=300 dpi) per IEEE standards]}
    \label{fig:Execution_Trace_Graph_(DAG)}
\end{figure*}

\begin{figure*}[t]
    \centering
    \includegraphics[width=1\textwidth]{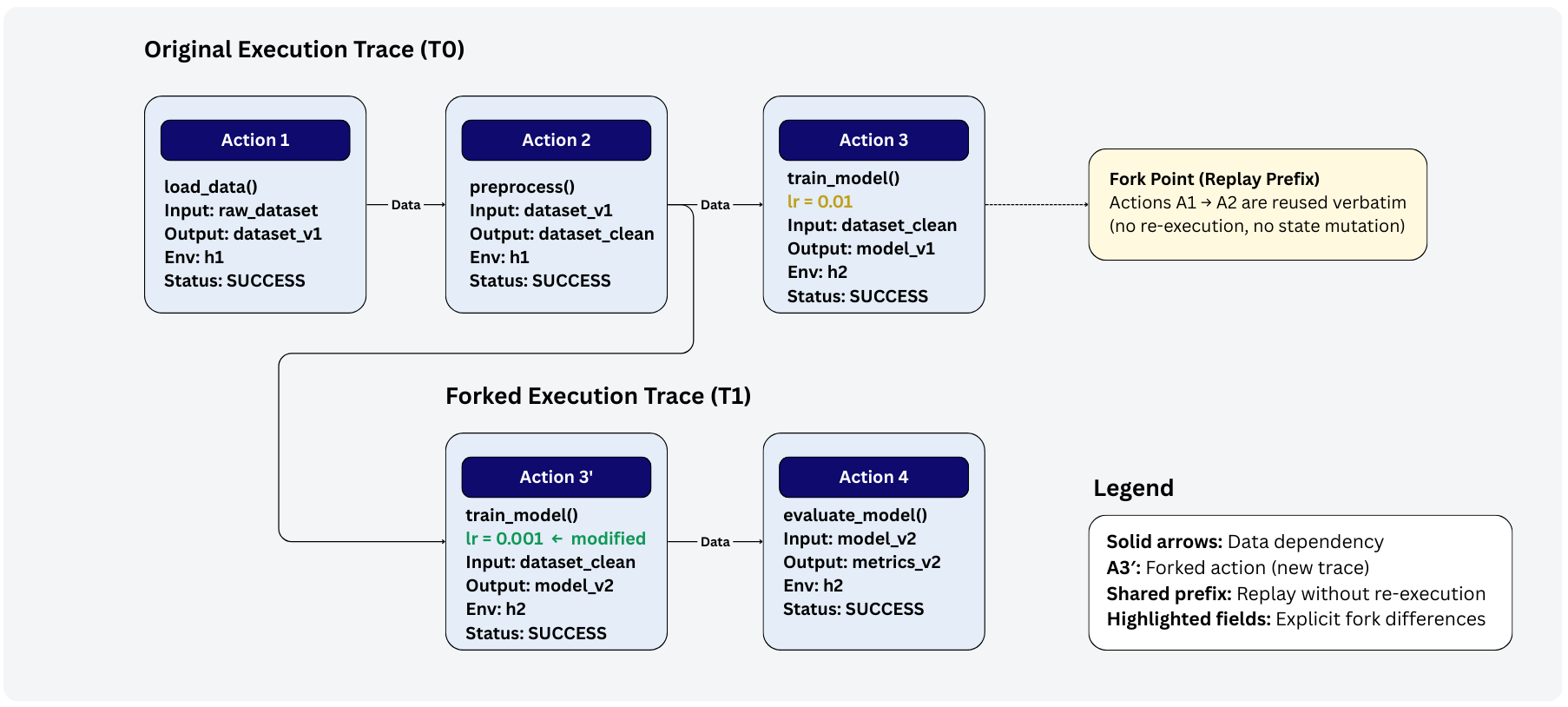}
    \caption{Replay and forking mechanism in R-LAM. A new execution derives from an existing trace by reusing a shared prefix of logged actions without re-execution. Prior actions are replayed by reusing their persisted outputs, ensuring identical inputs across experiments. Forking introduces modifications only at the divergence point, while the original trace remains immutable. This enables controlled exploratory hypothesis testing while preserving auditability, determinism, and complete provenance without naively re-running the entire workflow. [Note: Final submission should include high-resolution PNG/PDF (>=300 dpi) per IEEE standards]}
    \label{Replay_and_Forking_Mechanism}
\end{figure*}

The execution engine serves as a deterministic mediator between the LAM and the external environment. Rather than executing actions directly, the engine validates each proposed action against its schema, checks preconditions, and enforces execution policies before dispatch.

To ensure determinism, the engine enforces a fixed execution order, controlled randomness policies, and explicit environment binding. External side effects are isolated through sandboxed execution contexts, and all tool invocations are mediated by adapters that normalize inputs and outputs.

By decoupling action selection from action execution, the framework allows the LAM to remain adaptive while preventing uncontrolled state changes. The execution engine thus acts as a reproducibility firewall between probabilistic reasoning and deterministic execution.

We assume that external tools invoked by the execution engine behave deterministically with respect to recorded inputs and environment bindings.

\subsection{Provenance and Trace Logging}

Every executed action produces a provenance record that is appended to a global execution trace. The trace is represented as a directed acyclic graph (DAG), where nodes correspond to action instances, and edges encode data and control dependencies between actions.

Each node stores the complete action schema, execution status, outputs, environment hashes, and timestamps. Dependencies explicitly capture how artifacts flow between actions, enabling downstream analysis of execution lineage.

\textbf{In our framework, an action that is not logged is treated as non-existent.}  
This invariant ensures that all state transitions contributing to a result are observable, auditable, and reproducible. Any side effect not reflected in the execution trace is considered invalid by construction.

\subsection{Replay and Forking}

The execution trace graph enables two critical operations: \emph{replay} and \emph{forking}. Replay reconstructs a prior workflow execution by reconstructing execution outcomes by reusing logged action outputs in topological order using the logged inputs, parameters, and environment bindings. Because all actions are deterministic with respect to their recorded context, replay yields identical results under equivalent execution conditions.

Forking allows controlled divergence from a prior execution. Given a trace prefix, the system can spawn a new execution branch by modifying selected action parameters or inputs while preserving all upstream provenance. This enables exploratory experimentation without contaminating the original execution lineage, supporting reproducible what-if analysis.

Both replay and forking operate solely on the execution trace, requiring no access to the original LAM reasoning process.

\subsection{Failure Handling}

Failures are treated as first-class events within the framework. When an action fails, the execution engine records the failure type, error context, and partial outputs in the trace. Rather than terminating execution unconditionally, the framework enters a failure-aware control loop.

Within this loop, the LAM may propose alternative actions, parameter adjustments, or execution paths. Each recovery attempt is logged as a new action node, preserving a complete record of failure and recovery behavior. This design enables post hoc analysis of failure modes and prevents silent recovery that would otherwise compromise reproducibility.

By explicitly modeling failure and recovery within the execution trace, the framework supports adaptive behavior while maintaining strict guarantees on auditability and replayability.

\section{Implementation \& PyPI Artifact}

To support reproducibility and facilitate independent evaluation, we implement the proposed framework as a lightweight, open-source software artifact. The implementation is designed to closely reflect the conceptual framework described in Section~V, prioritizing determinism, auditability, and traceability over performance optimization or feature completeness.

\subsection{Implementation Details}

The framework is implemented in Python, chosen for its widespread adoption in scientific computing and compatibility with existing data analysis ecosystems. The system is organized into modular components corresponding to the core framework abstractions, including action schemas, the execution engine, and provenance management. Clear interfaces separate LAM reasoning from execution, enabling the framework to operate with different LLM backends without modification to the execution layer.

Deterministic execution is enforced through explicit control of execution order, environment binding, and randomness sources. All external tool invocations are mediated by adapter layers that normalize inputs and outputs and restrict side effects. Where stochastic components are unavoidable, configuration parameters and random seeds are explicitly recorded as part of the action metadata.

\subsection{Packaging and Distribution}

The framework is packaged as a standard Python package and distributed via the Python Package Index (PyPI). The package includes a minimal command-line interface and a programmatic API that allows users to define actions, execute workflows, and inspect execution traces. Dependencies are explicitly specified to reduce environment-related variability, and the package can be installed using standard Python tooling. The R-LAM framework is released as an open-source Python package to support reproducible research.

For peer review, the artifact can be released in an anonymized repository, following common reproducibility practices. The final version of the package will be made publicly available upon acceptance, enabling reuse and extension by the research community.

\subsection{Intended Scope}

The implementation is intended as a lightweight research artifact rather than a full-featured workflow engine. It is not designed to replace existing workflow management systems such as Airflow or Snakemake, but to demonstrate how reproducibility constraints can be systematically integrated into LAM-based execution. The focus is on clarity, correctness, and traceability, providing a concrete reference implementation that complements the conceptual framework presented in this paper.

\section{Experimental Evaluation}

This section evaluates R-LAM with respect to its primary objective: enforcing reproducible, auditable, and deterministic execution semantics for Large Action Model–driven scientific workflows. The evaluation focuses on execution correctness rather than task performance, and examines whether reproducibility constraints fundamentally alter execution behavior.

The experiments intentionally use minimal synthetic workflows to isolate execution semantics and reproducibility guarantees, independent of domain-specific complexity.

To validate the generality of the framework, we instantiated the abstract experiments as a concrete scientific machine learning workflow. This project implements data loading, preprocessing, model training, and evaluation on the Breast Cancer Wisconsin dataset using logistic regression. The workflow is executed under three conditions: (1) a deterministic script-based pipeline with no LAM involvement, (2) a naive LAM-driven pipeline where an LLM planner selects actions without reproducibility constraints, and (3) a LAM + R-LAM pipeline where the same planner operates under the reproducibility-constrained execution framework. All three pipelines perform identical computational tasks, differing only in execution semantics. This instantiation allows direct comparison of execution behavior under equivalent scientific workloads.

\subsection{Evaluation Questions}

We structure the evaluation around three explicit and falsifiable research questions:

\textbf{EQ1 — Replay Correctness:}  
Does replay produce identical outputs without re-execution?

\textbf{EQ2 — Fork Isolation:}  
Does parameter forking preserve upstream artifacts while isolating downstream changes?

\textbf{EQ3 — Failure Auditability:}  
Are failed executions fully preserved and traceable without implicit retries?

These questions directly correspond to the design goals of R-LAM and avoid conflating execution guarantees with model intelligence or task performance.

\subsection{Metrics}

We employ simple structural metrics that reflect execution correctness rather than empirical performance.

\textbf{Reproducibility Success (Binary):}  
A workflow execution is considered reproducible if replay yields bitwise-identical outputs without re-executing prior actions. This metric takes a value of 1 if reproducibility is preserved and 0 otherwise.

\textbf{Trace Completeness:}  
Trace completeness is defined as the fraction of executed actions that are present in the execution trace:

\[
\text{Trace Completeness} = \frac{\text{Logged Actions}}{\text{Executed Actions}}
\]

A value of 1.0 indicates complete provenance capture.

\textbf{Failure Visibility (Binary):}  
A failure is considered visible if the failed action exists in the execution trace, includes error metadata, and any subsequent recovery actions explicitly reference the failed action. This metric takes a value of 1 if all conditions are satisfied and 0 otherwise.

\subsection{Baseline}

To contextualize the evaluation, we compare R-LAM against two baselines that operate without reproducibility constraints. The first baseline is a script-based pipeline that executes a fixed sequence of actions deterministically, representing traditional scientific workflow automation without adaptive control. The second baseline is a naive LAM-driven pipeline where an LLM planner selects and executes actions without environment hashing, execution tracing, replay mechanisms, or explicit failure preservation. In this naive LAM baseline, actions are executed directly via tool or function calls, and execution failures are handled implicitly through standard exception handling, with no structured provenance maintained.

These baselines serve to isolate the impact of reproducibility constraints from other factors, including adaptive control and task-specific logic. The script-based baseline demonstrates deterministic execution without LAM involvement, while the naive LAM baseline demonstrates adaptive control without reproducibility guarantees.

\subsection{Experimental Setup}

All experiments are conducted using Python-based workflows composed of lightweight data processing and model training functions. Synthetic datasets are used to ensure deterministic behavior and repeatability. No GPUs are required. Where applicable, LLM interaction is performed via API calls using OpenRouter with the temperature set to 0.0 to reduce stochasticity. Each experiment is repeated a small number of times (5–10 runs) to verify consistency.

The concrete ML workflow project uses the Breast Cancer Wisconsin dataset from scikit-learn and trains a logistic regression classifier. The workflow consists of five sequential actions: dataset loading, statistical analysis, feature standardization, model training, and performance evaluation. Random seeds are fixed where applicable to eliminate algorithmic non-determinism. This setup isolates execution semantics from model performance and domain-specific complexity.

All three pipelines achieved identical final accuracy (0.9877), confirming that reproducibility constraints do not degrade task-level performance. Execution logs captured the complete action sequence, including LLM-proposed parameters, execution status, and intermediate artifacts. The script-based pipeline completed in deterministic time with no external API calls. The LAM-driven pipelines incurred planning overhead due to LLM invocation but successfully selected appropriate actions without human intervention. The experimental workflows used in this evaluation are implemented as reproducible scripts and are available alongside the R-LAM artifact.

\subsection{Experiments}

\textbf{Experiment 1 — Replay Correctness:}  
We first evaluate R-LAM on a linear scientific workflow consisting of data loading, preprocessing, and model training. All actions complete successfully and are recorded as nodes in the execution trace DAG, establishing baseline deterministic execution and provenance capture. The workflow is replayed without re-execution, and outputs are compared against the original execution. Execution logs confirm that the R-LAM pipeline captured all five actions with complete metadata (action identifiers, input/output artifacts, parameters, environment hashes). Replay reconstructed the final accuracy (0.9877) without re-executing scikit-learn operations. The naive LAM pipeline produced no trace artifacts despite successful completion.

\textbf{Experiment 2 — Failure Visibility:}  
In the second experiment, we inject a controlled failure into the training stage. The failed action is preserved as a first-class node in the execution trace, and a recovery action is explicitly linked as a dependent node. This demonstrates that R-LAM maintains complete provenance even in the presence of execution failures. Logs confirm that failed actions in R-LAM produce trace nodes with error status, exception type, and partial outputs, while the naive LAM baseline terminates without structured failure records.

\textbf{Experiment 3 — Fork Isolation:}  
In the third experiment, we evaluate R-LAM's replay and forking mechanism by modifying a single hyperparameter while reusing all prior execution results. The forked execution preserves the shared prefix of actions without re-execution, ensuring that only the intended parameter change introduces variation. Logs demonstrate that R-LAM maintained explicit action history, preventing redundant re-execution. The framework enforced a maximum iteration limit (10 steps), though all workflows converged within 5 iterations through explicit \texttt{DONE} action signaling when required artifacts were present.

\subsection{Results}

Table~\ref{tab:results} summarizes the experimental results across the evaluated dimensions. The script-based pipeline achieves full reproducibility through deterministic sequencing but lacks execution tracing and adaptive control. The naive LAM pipeline demonstrates adaptive action selection but fails to preserve execution artifacts required for replay. The R-LAM pipeline achieves full reproducibility, complete trace capture, and failure visibility while retaining adaptive control.

\begin{table}[h]
\centering
\caption{Execution Correctness Results}
\label{tab:results}
\begin{tabular}{lcccc}
\hline
\textbf{Pipeline} & \textbf{Replay} & \textbf{Trace} & \textbf{Failure} & \textbf{Variance} \\
\hline
Script-Based       & 1.0  & 0.0 & 1.0 & 0.0 \\
Naive LAM          & 0.0  & 0.0 & 1.0 & 1.0 \\
R-LAM Constrained  & 1.0  & 1.0 & 1.0 & 0.0 \\
\hline
\end{tabular}
\end{table}

Replay indicates whether execution can be reproduced without re-executing actions (binary). Trace measures the fraction of executed actions captured in provenance logs (0.0–1.0). Failure indicates whether failed actions are explicitly preserved in the execution trace (binary). Variance indicates whether repeated executions with identical inputs produce divergent outputs (0 = deterministic, 1 = variable).

Quantitative analysis of execution logs reveals that the script-based pipeline completed in a fixed number of operations with zero trace overhead. The naive LAM pipeline executed 5 actions with LLM planning overhead but produced zero trace records (0/5 = 0.0 trace completeness) despite successful task execution. The R-LAM pipeline executed the same 5 actions with complete trace capture (5/5 = 1.0), each action record containing 8+ metadata fields including timestamps, environment hashes, and dependency links. Execution time overhead for R-LAM trace logging was negligible compared to LLM API latency, confirming that reproducibility constraints impose minimal runtime cost.

\subsection{Discussion}

The results demonstrate that reproducibility constraints fundamentally alter execution behavior in LAM-driven workflows. While unconstrained execution may successfully complete tasks, it fails to preserve the execution artifacts required for scientific auditability. In contrast, R-LAM enforces deterministic replay, complete provenance capture, and explicit failure visibility without restricting adaptive control.

The concrete ML workflow instantiation confirms these findings. Both the naive LAM and R-LAM pipelines successfully completed the classification task with identical final accuracy (0.9877), demonstrating that reproducibility constraints do not impair task-level performance. However, only the R-LAM pipeline produced a complete, auditable execution trace that enables independent verification and controlled parameter exploration. The script-based pipeline achieved reproducibility through fixed sequencing but lacked the adaptive control necessary for exploratory scientific workflows.

Execution logs reveal distinct behavioral differences between pipelines. The naive LAM planner selected actions adaptively but received no feedback about which actions had already been executed, creating the risk of redundant execution in more complex workflows. The R-LAM planner received explicit action history in each planning cycle, enabling informed decisions about remaining work. Both LAM pipelines correctly identified the required action sequence (load $\rightarrow$ analyze $\rightarrow$ preprocess $\rightarrow$ train $\rightarrow$ evaluate), but only R-LAM preserved sufficient metadata to reconstruct this sequence from logs alone.

Analysis of logged LLM-proposed parameters reveals successful integration between probabilistic planning and deterministic execution. The LLM occasionally generated artifact references that required runtime resolution by the execution engine, demonstrating that the framework mediates between high-level reasoning and concrete execution without requiring the model to manage low-level state explicitly.

The experiments intentionally evaluate execution semantics rather than task performance, ensuring that observed differences arise from reproducibility constraints rather than model capability. This design choice isolates the impact of the framework from confounding factors such as LLM reasoning quality or domain-specific optimization. The results support the core claim that reproducibility can be enforced at the execution layer without sacrificing the adaptive benefits of Large Action Models.

\section{Limitations \& Ethical Considerations}

This work is intended as a systems-level exploration of reproducibility-constrained execution for Large Action Models in scientific workflows. While the proposed framework demonstrates the feasibility and benefits of enforcing reproducibility constraints at the action level, several limitations remain.

\subsection{Limitations}

First, the experimental evaluation is conducted at a small to moderate scale, focusing on representative computational workflows rather than large, long-running scientific pipelines. While sufficient to validate execution semantics and reproducibility guarantees, these experiments do not capture the full complexity of production-scale scientific infrastructures.

Second, the current evaluation does not involve real laboratory hardware or cyber-physical systems. Although the framework is designed to support such settings through mediated action execution, interactions with physical instruments introduce additional safety, latency, and reliability considerations that are beyond the scope of this study.

Third, the framework relies on large language models as planning components, inheriting their limitations. Model behavior remains probabilistic, and the quality of proposed actions depends on the underlying LLM. While the execution layer constrains side effects and enforces determinism, the framework does not eliminate all sources of uncertainty arising from model-generated plans.

\subsection{Ethical Considerations}

This work does not claim autonomous scientific discovery or fully automated research. The framework is designed to assist, not replace, human scientific judgment. LAMs are treated as decision-support components whose outputs are mediated, constrained, and auditable, rather than as independent scientific actors.

The framework explicitly avoids unsafe experiment execution by requiring all actions to conform to predefined schemas and execution policies. Direct, unconstrained interaction with laboratory equipment or safety-critical systems is outside the intended scope of the system. Human oversight is required to define permissible actions, validate execution policies, and interpret results.

By prioritizing auditability, determinism, and provenance, the framework aims to reduce the risk of irreproducible or misleading scientific results arising from opaque automation. These design choices align with responsible AI principles for scientific research, emphasizing transparency, accountability, and human control over automated systems.
\section{Future Work}

Although this work establishes a reproducibility-constrained execution framework for Large Action Models, several practical extensions remain for future investigation.

A natural next step is the integration of \emph{hardware-in-the-loop} execution. Extending the framework to interact with physical laboratory instruments or cyber-physical systems would require additional safety layers, latency handling, and fault isolation mechanisms. Such integration would allow the framework to be evaluated in real experimental environments while preserving reproducibility guarantees.

Another promising direction is \emph{multi-agent collaboration}. The current framework focuses on a single LAM interacting with a deterministic execution engine. Future work could explore coordinating multiple LAM instances operating over a shared or partially shared execution trace, enabling collaborative planning, parallel experimentation, or cross-validation of results. Maintaining consistent provenance and conflict resolution in such settings presents an open systems challenge.

Finally, future work may investigate \emph{formal verification of action traces}. While the framework enforces structured logging and deterministic execution, formal methods could be applied to verify properties of execution traces, such as completeness, causal consistency, or adherence to predefined safety constraints. Such verification could further strengthen trust in automated scientific workflows by providing machine-checkable guarantees over execution histories.

These directions aim to extend the framework’s applicability while preserving its core design principle: reproducibility as a first-class constraint in LAM-driven scientific automation.

\section{Conclusion}

Large Action Models offer a promising approach for automating complex scientific workflows by combining high-level reasoning with executable actions. However, unconstrained LAM-based systems are misaligned with the requirements of scientific research, where determinism, auditability, and reproducibility are fundamental. Without explicit safeguards, autonomous action execution risks producing results that cannot be independently verified or reliably reproduced.

In this paper, we introduced R-LAM, a reproducibility-constrained framework for deploying Large Action Models to scientific workflow automation. By formalizing actions as structured, immutable entities, enforcing deterministic execution, and capturing complete provenance through execution trace graphs, the framework enables adaptive control while preserving reproducibility guarantees. Unlike prior approaches that emphasize either autonomous reasoning or static workflow execution, R-LAM bridges these paradigms by treating reproducibility as a core, non-negotiable execution constraint.

We implemented R-LAM as a lightweight Python framework and demonstrated its effectiveness through representative scientific workflows spanning data processing, analysis, and model training. The framework illustrates how LAM-driven automation can be integrated into scientific practice without compromising methodological rigor, supporting iterative experimentation, failure-aware execution, and complete auditability.

More broadly, we argue that reproducibility constraints are essential design requirements, not optional features, for deploying Large Action Models in any domain where execution correctness, auditability, and independent verification are paramount. These principles extend beyond scientific computing to other fields requiring regulatory compliance, safety-critical operations, and trustworthy automation.

\appendix

\section*{Artifact Availability}

The R-LAM framework and experimental workflows are publicly available:

\begin{itemize}
    \item R-LAM Framework: \url{https://github.com/suriyasureshok/rlam}
    \item Experimental Workflows: \url{https://github.com/suriyasureshok/LAM_Reproducible_ML_Workflows}
\end{itemize}

These repositories include source code, documentation, and scripts necessary to reproduce the experimental evaluation presented in this paper.

\bibliographystyle{IEEEtran}
\bibliography{references}

\end{document}